\documentclass{article}

\usepackage{arxiv}

\usepackage[utf8]{inputenc} 
\usepackage[T1]{fontenc}    
\usepackage{hyperref}       
\usepackage{url}            
\usepackage{booktabs}       
\usepackage{amsfonts}       
\usepackage{nicefrac}       
\usepackage{microtype}      
\usepackage{lipsum}		    
\usepackage{hyphenat}       
\usepackage{graphicx}
\usepackage{natbib}
\usepackage{doi}

\title{Toward a Fully Autonomous, AI-Native Particle Accelerator}

\date{\vspace{-2em}February 12, 2026}

\author{ \href{https://orcid.org/0000-0003-3814-8417}{\includegraphics[scale=0.06]{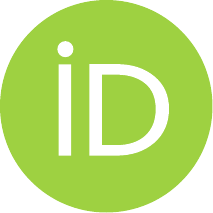}\hspace{1mm}Chris ~Tennant} \\
	Jefferson Laboratory\\
	12000 Jefferson Ave\\
	Newport News, VA 23606 \\
	\texttt{tennant@jlab.org} \\
}

\renewcommand{\headeright}{}
\renewcommand{\undertitle}{}

\hypersetup{
pdftitle={Toward a Fully Autonomous, AI-Native Particle Accelerator},
pdfsubject={},
pdfauthor={Chris ~Tennant},
pdfkeywords={autonomous, self-driving, particle accelerator, AI, agentic},
}

\begin{document}
\maketitle

\begin{abstract}
This position paper presents a vision for self-driving particle accelerators that operate autonomously with minimal human intervention. We propose that future facilities be designed through artificial intelligence (AI) co-design, where AI jointly optimizes the accelerator lattice, diagnostics, and science application from inception to maximize performance while enabling autonomous operation. Rather than retrofitting AI onto human-centric systems, we envision facilities designed from the ground up as AI-native platforms. We outline nine critical research thrusts spanning agentic control architectures, knowledge integration, adaptive learning, digital twins, health monitoring, safety frameworks, modular hardware design, multimodal data fusion, and cross-domain collaboration. This roadmap aims to guide the accelerator community toward a future where AI-driven design and operation deliver unprecedented science output and reliability.
\end{abstract}

\keywords{autonomous, self-driving, particle accelerator, AI, agentic}

\section{Introduction}
Imagine a particle accelerator that runs itself, automatically tuning thousands of magnets and RF cavities for optimal beam quality, detecting and diagnosing equipment faults before they cause downtime, and adapting in real-time to changing experimental demands -– all with minimal human intervention. As accelerators grow even more powerful and intricate, with millions of sensor channels and thousands of interconnected components that must be precisely coordinated, human operators will be stretched to their limits. This has prompted a fundamental question: can we operate accelerators more autonomously, with AI managing complexity at machine speed while humans provide strategic oversight?

The vision of a “self-driving, natively-AI” particle accelerator is a facility that runs optimally with minimal human intervention, continuously tuning itself, diagnosing issues, and safely adapting to changing conditions in real time. Realizing this vision requires rethinking both control systems and accelerator design from the ground up. An autonomous accelerator needs AI systems deeply integrated at every level, from low-level device control up to high-level decision-making. Crucially, the accelerator must be natively AI-driven, meaning it is engineered for autonomy from the outset rather than having AI added as an afterthought. Every subsystem should be fully instrumented, digitally accessible, and outfitted with automation hooks for AI to auto-configure, auto-stabilize, auto-analyze, and auto-recover as needed. Like a driverless car’s sensor fusion and autopilot layers \cite{01_Kocic2018}, future accelerators will embed intelligent diagnostics and control loops throughout for safe and efficient operation. The following sections outline this vision in more detail and draw connections to current advances that hint at what is possible \cite{02_Edelen2024}. This vision aligns directly with recent national priorities, as evidenced by the Department of Energy's designation of "Enhancing Particle Accelerators for Discovery" as one of 26 Genesis Mission Science and Technology Challenges, specifically calling for deploying AI to make accelerators adaptive and autonomous to accelerate breakthroughs in medicine, materials, and energy \cite{DOE2026Genesis}.

\section{A New Paradigm for Accelerators: AI-Driven Design and Operations}
Enabling a truly autonomous accelerator calls for a paradigm shift in how we design and operate these machines. In this paradigm, AI agents serve as the primary operators of the facility, while humans shift to supervisory and strategic roles. The accelerator complex would effectively become a cyber-physical AI system. Every cavity, magnet, diagnostic, and power supply is not only digitally controlled but also paired with AI routines that monitor and adjust its performance. On-site human intervention would become a last resort. Achieving this means engineering accelerators with autonomy as a core design goal, incorporating built-in safety margins, high component reliability, modular designs, and redundancy so that automated adjustments or component swaps can occur without jeopardizing operations.

\subsection{AI Co-Design: Optimizing the Accelerator and Science Application Together}
\label{subsec:codesign}
The vision of an autonomous, natively-AI accelerator begins not with operations, but with design. A truly autonomous facility must be conceived through AI co-design from its inception, where AI shapes both the accelerator architecture and its science application as a unified, jointly-optimized system. This represents a fundamental departure from conventional approaches where accelerator design and experimental requirements are treated as separate optimization problems, connected only through interface specifications and performance requirements.

AI-driven design would leverage accelerator-specific knowledge bases or foundation models to probe vast solution spaces and identify novel lattice configurations that go beyond current state-of-the-art performance \cite{03_Ji2024}. Rather than relying solely on human intuition and incremental improvements to established designs, AI systems can explore unconventional combinations of components to discover configurations that beam physicists might never consider. Crucially, this optimization extends beyond the traditional accelerator components to include the diagnostics themselves –- which sensors to deploy, where to place them, and how to configure them for optimal self-diagnosing capabilities. An AI-designed accelerator would thus be born with the instrumentation and observability required for autonomous operation, rather than having diagnostics added as an afterthought.

More significantly, the science application –- whether it be a nuclear physics experiment or a materials characterization beamline –- would be jointly optimized with the accelerator lattice from the very beginning. AI co-design enables optimization of the entire system – accelerator and experiment together –- toward maximizing specific science objectives. This integrated approach promises to generate outsized science output compared to conventionally designed facilities. We envision demonstrating massive improvements in metrics such as science-output-per-unit-time, science-output-per-dollar, or science-output-per-square-foot, depending on what constraints matter most for a given application.

Such improvements become possible because AI design can simultaneously optimize across multiple, often competing constraints: physical footprint, energy and power consumption, capital and operational costs, and reliability. Consider reliability as an example. Traditional design processes rely on conservative safety factors and component specifications, but an AI system could leverage historical operational data from specific components – manufacturer, model, and configuration – deployed across multiple laboratories worldwide. By learning which designs and component choices have proven most reliable in practice, and under what operating conditions they degrade or fail, the AI can make informed trade-offs that balance performance against long-term dependability. A design optimized for reliability from the start will be far easier to operate autonomously, as the system experiences fewer unexpected failures and operates more consistently within well-characterized regimes.

The paradigm shift here is profound, rather than designing an accelerator and then attempting to automate its operation, or designing an experiment and then building an accelerator to serve it, we design the entire integrated system – accelerator, diagnostics, control architecture, and science application – as a unified whole optimized by AI for both peak performance and autonomous operation. This is what we mean by a “natively-AI” accelerator. The autonomy is not retrofitted, rather it is intrinsic to the machine’s very architecture because that architecture was conceived with autonomy as a core design objective.

\subsection{Staged AI Integration: From Assistance to Autonomy}
To contextualize this vision, it helps to frame the progression as three stages of AI integration in accelerator operations. Current facilities operate in the \textit{AI-assisted} stage, where AI tools are retrofitted onto human-centric systems to help operators, but humans remain firmly in control. Near-term developments will move us toward the \textit{AI-augmented} stage, where facilities are designed to seamlessly integrate AI tools and enable collaborative human-AI operation. The ultimate goal described in this paper is the \textit{AI-autonomous} stage, where facilities are designed from inception with AI as the primary operator and humans provide oversight and strategic guidance. While existing facilities will progress through these stages evolutionarily -— advancing from AI-assisted to AI-augmented through incremental upgrades and operational changes -— the AI-autonomous stage will be most fully realized in greenfield facilities purpose-built for autonomy from the ground up. These new facilities will inherit lessons learned from the evolutionary path while avoiding the constraints inherent in retrofitting autonomy onto legacy systems. This staged framework makes explicit that the autonomous accelerator is the natural endpoint of a trajectory already underway, building on foundations being laid today.

\subsection{Natural Language Interfaces and Human-AI Collaboration}
A natively-AI accelerator also implies a fundamentally different human-machine interface \cite{04_Hellert2025}. In an AI-autonomous facility, the interface serves not for routine operational control, but for strategic oversight, intervention, and understanding. Recent work on large language model (LLM) based assistants for beamlines demonstrates the enabling technology for this supervisory paradigm \cite{05_Mayet2024, 06_Hellert2025, 07_Hellert2025}. Rather than today's low-level control panels where operators manually adjust thousands of parameters, the interface allows humans to monitor what the AI is doing, understand why it made particular decisions, set high-level objectives, and intervene when necessary. For example, a supervisor might query "why did you reduce the beam current in Hall A this morning?" and receive a natural language explanation grounded in sensor data and the AI's reasoning. Or they might establish a new strategic directive: "prioritize Hall B experiments for the next 48 hours while maintaining Hall A in standby." The AI would then autonomously reconfigure the machine to meet these goals. In edge cases or novel situations where the AI lacks confidence, it would present options in natural language: "I've detected an anomaly in the cryogenic system. I can either reduce beam power by 20\% immediately, or attempt controlled shutdown of the affected module. What do you prefer?" This mode of interaction creates a partnership where AI handles moment-to-moment complexity at machine speed while humans provide strategic direction, validate major decisions, and serve as the ultimate authority for safety-critical choices. The interface makes the autonomous system's behavior transparent and its decision-making process inspectable, which is essential for building trust and maintaining human oversight even as direct operational control recedes.

\section{Current Progress and Emerging Capabilities}

Over the last several years, the accelerator community’s engagement with AI has expanded rapidly – from early demonstrations to a broad, community-wide push spanning modeling, tuning, diagnostics, and reliability.

A major thrust has been AI-enhanced accelerator modeling, where learned surrogate models complement physics-based simulation by providing fast, high-fidelity predictions that can be used in both design studies and operations \cite{09_Edelen2020, 11_Arpaia2021}. In many contexts, these learned models can run orders of magnitude faster than full simulations while preserving sufficient accuracy to support online prediction, calibration, and control-oriented use cases. Alongside this, there is growing emphasis on hybrid physics-informed approaches \cite{12_Ivanov2020}, model calibration/adaptation, and efficient sampling strategies that help keep models trustworthy as machines drift and operating conditions change. 

In parallel, there has been strong momentum in online optimization and control, where AI methods augment or replace manual tuning by searching high-dimensional parameter spaces more efficiently and with explicit attention to operational constraints. This includes Bayesian and model-guided optimization \cite{13_Roussel2024}, learned online models \cite{14_Leemann2019}, and increasing interest in reinforcement learning (RL)--style adaptive control policies \cite{16_Kain2020, 17_Hirlaender2023, 18_Kaiser2022} – often coupled to the broader ecosystem of simulations and digital twins for safe development and validation. Recent work on adaptive, real-time triggering systems at the LHC demonstrates RL agents responding to changing experimental conditions autonomously \cite{selfdrivingtriggerlhc}. Beyond tuning, AI is also increasingly used for design optimization \cite{19_Zhang2025}, AI-enhanced diagnostics and virtual diagnostics \cite{20_Scheinker2025}, and anomaly detection / failure prediction \cite{21_Tennant2020, 22_Blokland2022}. Beyond algorithms for tuning and control, more agentic AI frameworks are being explored to coordinate higher-level operations \cite{23_Sulc2025, 06_Hellert2025}.

All of these developments provide valuable testbeds. They validate elements of the autonomous accelerator concept (albeit in isolation), and they highlight the research frontiers that must be addressed to integrate these pieces into a cohesive, autonomous whole. With the recent launch of the Genesis Mission, these community-wide efforts will be further accelerated through coordinated national investment \cite{24_trump2025}. These efforts align with strategic priorities identified in both the U.S. Accelerator and Beam Physics Roadmap \cite{ABPRoadmap2023} and the European Strategy for Particle Physics Accelerator R\&D Roadmap \cite{ESPPAccelRoadmap2024}, which recognize AI and machine learning as transformative capabilities for future facilities.

\section{Major AI Research Thrusts Required to Realize the Vision}
\begin{figure}
    \centering
    \includegraphics[width=1\linewidth]{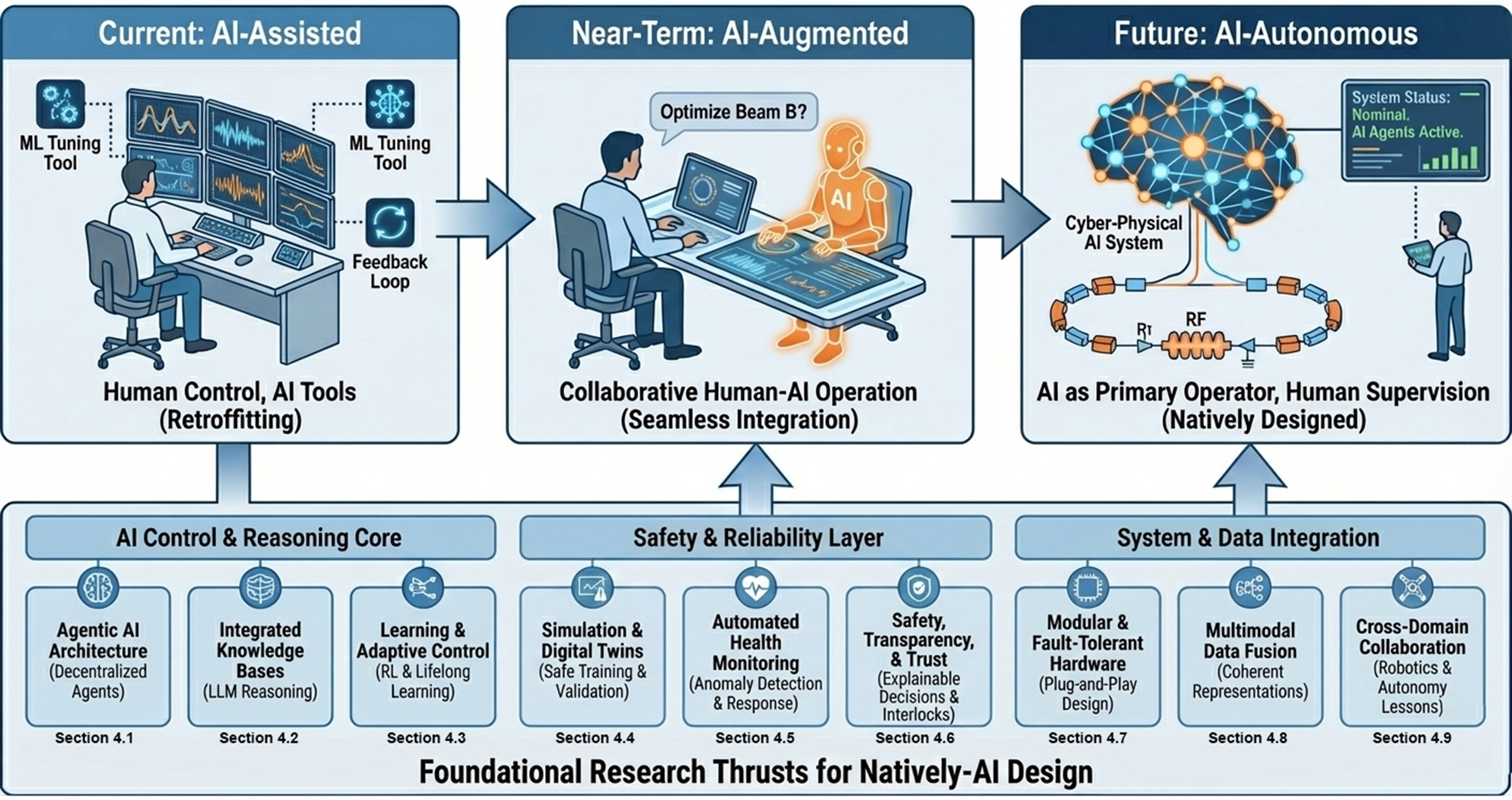}
    \caption{Illustration of the stages of AI progress in accelerator operations (top) and the nine research thrusts needed to achieve autonomous operation (bottom).}
    \label{fig:fig1}
\end{figure}

Transitioning from isolated successes to an autonomous, natively-AI accelerator will require sustained research and development across multiple areas of AI, accelerator science, and systems engineering. We outline below the major thrusts that must converge to achieve this vision. Each thrust represents a critical capability or framework that is currently nascent or missing in accelerator operations. These research areas span from the software and algorithmic core of AI agents to the hardware and architecture of the accelerator itself. Together, they form a roadmap for transforming accelerator facilities into autonomous, learning machines. See Fig. \ref{fig:fig1}.

\subsection{Agentic AI Control Architecture}
\label{subsec:agent}
A foundational requirement is an AI-driven control architecture that can manage the accelerator’s complexity through coordinated agents. Rather than monolithic software, the system should consist of specialized AI agents, each responsible for different subsystems or tasks, that communicate and collaborate to run the machine. Recent proposals advocate a decentralized multi-agent approach in which high-level cognitive agents handle reasoning and communication, while lower-level agents execute domain-specific control actions \cite{23_Sulc2025}. For example, one agent might monitor and optimize the RF cavity performance, while another oversees beam orbit corrections. A higher-level supervisor agent could synthesize their reports and make global decisions. This modular, agent-based architecture promises to be more robust and scalable than a single AI agent, reflecting the way human expert teams operate accelerators in parallel. Crucially, LLMs have emerged as a key enabling technology for agentic systems. They can serve as the “brains” of agents that need to interpret complex sensor data, read logs or documentation, and communicate in natural language with operators and other agents \cite{25_Schick2023}. Early experiments have shown LLM-driven agents capable of interpreting accelerator state and even generating operational plans or code snippets on the fly \cite{23_Sulc2025}. Designing an effective agent architecture will involve deciding how to distribute responsibilities among agents, what communication protocols they use, and how much autonomy each has \cite{26_Yao2023}. The ultimate goal is an intelligent orchestration of agents that collectively operate the accelerator.

Equally important is establishing the right balance between decentralized autonomy and centralized coordination. In an autonomous accelerator, we expect a hybrid model. Certain agents might act with local autonomy (tuning a magnet loop independently), but a coordinating mechanism must align their actions to avoid conflict or suboptimal outcomes. Multiple architectural approaches are possible. One approach is a central orchestrator that plans and approves agent actions. This is the pattern adopted by the Osprey framework, which uses a single planning agent to route tasks to specialized capabilities in a controlled, transparent way \cite{06_Hellert2025}. The orchestrator concept emphasizes safety and predictability. Before any action is taken, a detailed plan is formulated and checked against constraints (e.g. no hardware limits are violated). Alternatively, a more distributed approach might let agents negotiate or vote on actions, which could be more scalable or fault-tolerant. Research is needed to determine which architecture (or combination thereof) best meets the demands of accelerator operations. It will likely draw on insights from multi-agent systems research and existing control system hierarchies \cite{27_Hughes2025}. The end result should be a unified AI control system where each agent’s contributions are synchronized toward global machine goals.

\subsection{Integrated Knowledge Bases and Reasoning}
\label{subsec:knowledge}
Effective autonomous operation will require AI agents that possess a deep and contextual understanding of the accelerator. This means developing an extensive knowledge base about the machine's design, constraints, and operational history. The foundation for this knowledge base begins not with operations, but during the AI co-design phase discussed in Section \ref{subsec:codesign}. During design, AI systems can ingest and reason over troves of existing documentation, schematics, control system data, and expert knowledge from across the accelerator community. Large language models provide a starting point, having demonstrated the ability to serve as knowledge engines when fine-tuned or augmented with domain data \cite{28_Xie2025}. By learning from decades of accelerator designs, component choices, failure modes, and operational patterns across facilities worldwide, the AI can inform design decisions that optimize for reliability and autonomous operation. This historical knowledge helps answer questions like: which magnet configurations have proven most stable? What diagnostic placements enable fastest fault isolation? Which RF cavity designs minimize trip rates? The logbook and documentation agents envisioned in recent agentic AI proposals demonstrate this capability to extract actionable insights from unstructured historical data \cite{29_ARIEL}.

A particularly promising development is using the accelerator’s machine state itself as a modality for foundation models. Recent work on multi-modal machine-state embeddings demonstrates this approach \cite{30_Mayet2025}. The machine state can be represented as a structured list of control-system (address, value) tuples – thousands to tens of thousands of setpoints and readbacks – optionally enriched with run metadata. The core idea is to embed this structured state into the language model’s representation space and condition the LLM through a learned adapter (e.g., cross-attention over a compact set of latent state vectors). An operator can query the machine in natural language and receive state-informed answers. Conversely, in the same way that multimodal LLMs not only “see” images, but can generate images based on text descriptions, the model can be asked to propose physics-informed machine states consistent with a requested operating mode, constraints, or science goal. This represents a fundamentally new way to interface with accelerator control systems, treating the complete machine configuration as interpretable data rather than just parameter lists.

As the facility is built and commissioned, the AI's knowledge base evolves from design intent and historical learning to include the specific as-built configurations, commissioning data, and ultimately the facility's own operational experience. This continuity –- from design through construction to operation –- ensures that the autonomous system inherits a deep understanding of why this particular machine was built the way it was, what trade-offs were made, and what performance envelopes were intended. An AI that helped design the accelerator is uniquely positioned to operate it autonomously. Critically, the operational knowledge base for an AI-autonomous accelerator will be built from that machine's own experience, not transplanted from legacy facilities operating under different paradigms.

Moving forward, an autonomous accelerator’s AI should maintain an ever-growing collective memory of operational experiences. Each incident, tuning adjustment, or anomaly encounter is a valuable lesson. By capturing these in a structured way (for instance, storing situations and successful responses), the AI agents can avoid repeating mistakes and improve their strategies over time. Such agents could, for instance, infer that a pattern of beam loss correlates with a temperature drift in a magnet, and then suggest that as a hypothesis to test. Achieving this will likely leverage a combination of LLM-based reasoning (to connect disparate pieces of information) and traditional physics knowledge encoded in the system. Another aspect is bridging the gap between human and machine representations, translating a high-level intent or diagnostic question into the low-level control language. All told, building a rich knowledge integration and reasoning capability into the AI operator is essential for autonomy. It ensures the AI’s decisions are informed by the full context of the machine and past expertise, rather than being myopic or purely reactive.

\subsection{Learning and Adaptive Control}
\label{subsec:learning}
An autonomous accelerator must be able to learn from experience and improve its control policies over time. Unlike fixed algorithms, the AI agents should adapt to changes in the machine and environment, learning to retune more efficiently as a magnet ages or as new beam modes are introduced \cite{31_Scheinker2013}. Reinforcement learning offers a framework for this kind of adaptive control \cite{32_Sutton}. An RL agent continuously adjusts control knobs (like magnet currents or RF phases) and receives feedback (such as beam quality metrics), iteratively learning strategies that maximize performance. The training strategy combines offline learning in digital twins with online refinement during operations. Initial RL policies would be trained extensively in simulation, where the agent can safely explore the full state space and learn robust control strategies without risking the physical machine (see Section \ref{subsec:digitaltwin}). Once deployed, these policies continue to adapt and improve through carefully constrained online learning, refining their performance based on real operational experience while respecting strict safety boundaries. Reinforcement learning has already shown promise in accelerator applications \cite{33_Kain2020, 34_Pang2020, 35_eichler2021}. However, applying RL at the scale of a whole accelerator raises open questions. One major challenge is balancing exploration with safety during online refinement \cite{36_garcia2015}. How can an RL agent safely try new control actions on a live, safety-critical machine while still learning and improving? The field of robotics has grappled with this by using simulations and digital twins for initial training before deploying on physical robots \cite{37_Tobin2017}. Another challenge is the sheer complexity and time-delay of accelerator systems. Reward signals may depend on a long sequence of actions and external conditions, making the credit assignment difficult. Research will be needed in advanced RL algorithms that can handle long-horizon tasks, incorporate prior physics knowledge to guide learning, and maintain safety guarantees during online adaptation.

Beyond pure RL, continuous learning and adaptation mechanisms are required so that the AI doesn’t stagnate. The accelerator’s AI should be able to refine its models and strategies with every run, essentially performing lifelong learning. Recent work suggests this is feasible \cite{38_Rajput2025}. One proposal describes agents that gradually improve through experience, continuously learning from operational data and incorporating human feedback when available \cite{23_Sulc2025}. Imagine an agent that tries a new approach to optimize beam stability. If it results in even a marginal improvement, the system should record that outcome and update its policy. Conversely, if an action leads to a negative outcome, the AI needs to adjust to avoid repeating it. In short, the AI operator must not be static. It should behave more like a seasoned physicist who keeps learning new tricks, except it does so at machine speed. This adaptive learning thrust will ensure the autonomous accelerator actually gets smarter and more efficient with time.

\subsection{Simulation and Digital Twins for Safe Autonomy}
\label{subsec:digitaltwin}
Because particle accelerators are high-consequence environments, simulation will be an indispensable tool for developing and validating AI control strategies. A detailed digital twin of the accelerator –- a real-time, faithful software replica of the machine -– can serve as a training ground and sandbox for AI agents \cite{40_Miceli2025, 41_Liuzzo2023}. Before an AI agent tries a novel tuning method or an RL policy on the actual accelerator, it can be tested against the digital twin to ensure it behaves safely and effectively. This approach mirrors what is done in autonomous vehicles and robotics, where virtual environments are used extensively to train and evaluate AI controllers. The digital twin would model not only the accelerator’s physics (beam dynamics, magnet fields, RF cavities, etc.) but also system behavior like timing, faults, and sensor noise, providing a high-fidelity testbed. Differentiable simulations are an especially promising area \cite{42_Huhn2025, 43_Wan2025, 44_Kaiser2024, 45_Roussel2022, 46_Gonzalez2023, 47_Qiang2023}. By simulating the accelerator in a way that provides gradient information, AI algorithms can more efficiently optimize control actions. For example, a differentiable model of beam optics could allow an AI agent to compute how small changes in magnet settings will affect the beam, enabling rapid convergence to optimal settings without blind trial-and-error.

In addition to training AI, digital twins will be vital for scenario testing and risk mitigation. The autonomous system can practice handling rare failure scenarios in simulation that might be impractical or dangerous to provoke in the real machine (such as sudden magnet quench or an RF breakdown). This ensures the AI is prepared for edge cases and can respond gracefully rather than being caught off guard. Furthermore, during operation, the digital twin can run in parallel with the real accelerator, continuously receiving live data. The AI could use it to forecast near-future machine states or to validate actions by checking if the simulated outcome flags any issues before committing to the real hardware. One open question is how to maintain and synchronize the digital twin as the machine evolves. It may require automated system identification techniques so the simulation parameters stay aligned with reality. Overall, this research thrust aims to provide the safe playground and validation layer that an autonomous accelerator’s AI will rely on. By combining the best of accelerator physics simulation and modern AI training practices, we can ensure that learning and optimization occur in a risk-free manner and that only thoroughly vetted strategies make their way into the live machine.

\subsection{Automated Health Monitoring and Anomaly Response}
\label{subsec:health}
High reliability is a non-negotiable requirement for any large scientific facility, and an AI-driven accelerator must excel at monitoring its own health and handling anomalies. With thousands of sensors and devices, manual monitoring is infeasible. Already today, operators struggle to sift through torrents of data to catch early signs of trouble. The AI agents should therefore take on the role of vigilant diagnosticians. Drawing inspiration from AI coding agents that autonomously debug software -- using domain knowledge to generate hypotheses, run tests, interpret results, and apply fixes -- an autonomous accelerator would diagnose and recover from degraded performance through a similar systematic loop: continuously analyzing facility data, generating plausible fault hypotheses, planning and executing safe diagnostic interventions, interpreting outcomes, implementing corrective actions, and recording what worked so future responses become faster and more reliable. This involves developing algorithms for real-time anomaly detection that can pinpoint subtle deviations in sensor patterns or equipment performance. Modern approaches like unsupervised learning and neural networks can identify outliers in multi-dimensional data without explicit thresholds \cite{48_Humble2022, 49_Ferguson2025}.

We envision an expansion of this idea with AI monitors running on every important subsystem (magnets, RF stations, vacuum, cryogenics, etc.). At this stage, it remains an open question whether this requires dedicated hardware and compute for each subsystem or a centralized processing architecture. Both approaches have merits and will require further investigation to determine the optimal computational architecture for real-time monitoring at scale. A higher-level agent would fuse alerts from subsystem monitors with beam quality metrics to decide if an intervention is needed. Importantly, these AI monitors can leverage techniques like anomaly clustering to not only detect but also classify the type of fault, enabling faster and more precise root cause analysis.

Once an anomaly or incipient fault is detected, the autonomous system needs protocols for automated mitigation and recovery. This could range from gracefully ramping down a section of the machine, to adjusting beam parameters to work around the problem. In some cases, the AI might fix the issue entirely on its own (for example, by resetting a tripped RF cavity or re-tuning a beamline for lossless transmission). In others it will alert human experts with a detailed diagnosis and recommended action plan. The ultimate goal is predictive maintenance, fixing problems before they lead to failures. AI can contribute here by analyzing long-term trends and subtle shifts in equipment behavior that humans might miss \cite{50_Garza2024}. For instance, slight increases in a magnet’s coil temperature combined with fluctuating beam orbit readings could signal a cooling issue developing. An AI agent could flag this and schedule a check or adjust operating conditions proactively. Early deployments of such capabilities are already in motion \cite{51_Lobach2024}. Extending these to a full-facility scale, with the AI effectively acting as a 24/7 sentinel, will be a major thrust. It not only improves uptime and performance but also instills trust that an autonomous accelerator won’t drift into unsafe territory without immediate correction.

\subsection{Safety, Transparency, and Trustworthiness}
\label{subsec:safety}
Deploying AI for safety-critical control demands rigorous attention to safety and transparency. The autonomous accelerator’s decisions must be explainable to human operators and fail-safe mechanisms must be in place for when AI goes awry. One research thrust focuses on designing the AI control system such that it can be thoroughly audited and constrained. The Osprey framework is instructive here. It enforces a plan-first approach where the AI must produce a complete, human-readable execution plan (including any device commands) for operator review before executing actions. Dangerous actions (like those involving high-power hardware changes) can be automatically flagged and paused for explicit approval. This kind of transparent orchestration greatly increases trust, as operators are not asked to accept a black box decision. We foresee future AI control systems embedding similar features. Every autonomous adjustment the accelerator makes would be logged with the rationale, and if the plan deviates from expected safe parameters, built-in interlocks halt the AI’s actions. Essentially, the traditional machine protection systems and software interlocks remain in force, with AI working within those boundaries rather than replacing them.

Another aspect is guarding against the unique failure modes of AI, such as hallucinations or erroneous reasoning by LLM-based agents. In non-critical settings, an AI mistake might be a harmless curiosity, but in an accelerator control room it could lead to downtime or equipment damage. Research is needed on techniques to keep AI outputs grounded and reliable. Ideas include using constrained generation (restricting AI suggestions to a formal grammar of known-safe commands), deploying multiple agents to cross-verify each other’s conclusions, and incorporating real-time sensor feedback to double-check AI predictions against reality. For instance, if one agent determines a magnet’s field should be increased, a separate watchdog agent might simulate the outcome first or ensure that sensor readings corroborate the need. There is also active work on improving AI factual accuracy through structured debate between models, which could be adapted to have two subsystems confirm a critical action \cite{52_Du2023}. 

Moreover, the autonomous system must have robust fallback strategies. If an AI component crashes or produces uncertain outputs, control should automatically revert to proven, traditional control algorithms or to a safe standby mode. Continuous validation against human operator decisions and established baselines will be important, especially in the early phases of deployment. Ultimately, this thrust is about ensuring the AI is a transparent and reliable guardian of the accelerator, one that earns the trust of the scientists and engineers by behaving predictably, explaining itself, and aligning with the safety principles of the facility.

\subsection{Modular and Fault-Tolerant Hardware Design}
\label{subsec:hardware}
Achieving an AI-run accelerator isn’t only about software and algorithms. It also calls for reimagining the hardware and system design with autonomy in mind. Future accelerators must be built to facilitate automated control and quick recovery from faults. In other words, modularity and fault tolerance should be core principles. A major thrust is to design accelerator components and subsystems that are plug-and-play and have built-in redundancy. Modular designs mean that if one component fails or needs maintenance, an AI system can isolate and replace it (if a redundant module is available). The European Spallation Source (ESS) and other modern projects have been embracing modular architectures (for example, using an array of many moderate-power modulators instead of one or two high-power sources) to improve overall reliability and serviceability \cite{53_Garoby2017}. This trend directly supports autonomy. An AI can far more easily manage self-repair or reconfiguration when there are many interchangeable modules and clear interface standards, as opposed to monolithic, unique systems.

However, the challenge of achieving true plug-and-play modularity in particle accelerators should not be underestimated. We are dealing with systems that may require breaking vacuum, venting cryogenic systems, and maintaining incredibly tight alignment tolerances. A superconducting magnet or RF cavity cannot simply be swapped like a modular power supply in a data center. Significant engineering and design breakthroughs will be needed to make autonomous component replacement a reality. This might involve developing new approaches to vacuum coupling, cryogenic connections, and precision alignment that can be performed robotically or with minimal human intervention.

In addition to modularity, engineering for high reliability and predictable failure modes is essential. Autonomous operation demands that unexpected hardware behaviors are minimized, and when failures occur, they do so in known, manageable ways. That might mean incorporating more sensors for early warning, using conservative operating margins, and providing robotic access for repairs. Robotic inspection and maintenance devices could be integral to the facility, allowing the AI to dispatch a robot to swap a failed control board or fix a vacuum leak without a human having to intervene \cite{54_Gamper2024, 55_Gentile2023, 56_PradosSesmero2021, 57_Angrisani2018}. We can draw an analogy to modern data centers where redundancy and remote-servicing capabilities are designed in from the start. In fact, a fully autonomous accelerator would treat on-site human fixes as a last resort, aiming instead to have automated fixes that keep the science running. This thrust requires collaboration between AI experts and accelerator engineers to ensure the physical machine is AI-friendly. Every subsystem should answer the question: if this breaks at 3AM, can the AI either fix it or work around it? By incorporating that philosophy, we pave the way for near-continuous operation with minimal downtime, which is a key promise of an AI-operated facility.

\subsection{Multimodal Data Fusion}
\label{subsec:multimodal}
The facility remains inherently heterogeneous, with fast-sampled waveforms, images from diagnostic cameras, event streams, alarms, sequencer and configuration files, and structured metadata describing mode, user requirements, and provenance. An autonomous accelerator must fuse these diverse data types into coherent internal representations. Multimodal AI capabilities enable the system to compare predicted versus observed signatures across different modalities and detect deviations that matter \cite{57_Angrisani2018}. For instance, correlating thermal camera images of RF stations with power readings and beam quality metrics could reveal subtle degradation patterns invisible to single-modality monitoring.

However, several research challenges must be addressed to realize effective multimodal fusion at scale. How do we develop foundation models that can natively process and reason over heterogeneous accelerator data -- combining time-series sensor data, diagnostic images, event logs, and structured configuration files -— without requiring separate preprocessing pipelines for each modality? What neural architectures enable real-time fusion of high-bandwidth data streams (potentially gigabytes per second) while maintaining the low latency required for closed-loop control? How can we ensure that learned multimodal representations remain interpretable and grounded in physics, rather than becoming inscrutable black boxes? Additionally, multimodal fusion must be robust to missing or corrupted data streams, as sensor failures and communication dropouts are inevitable in large-scale facilities.

This fusion also supports a reimagined logbook for autonomy. The AI agent can generate human-readable, physics-grounded shift-style reports and machine health narratives as a byproduct of perception and control, rather than relying on inconsistent manual reporting. By treating the full spectrum of accelerator data as interpretable information, the autonomous system achieves a more complete situational awareness than is possible with conventional monitoring approaches. Developing the architectures, training methodologies, and real-time inference systems to enable this comprehensive multimodal understanding represents a critical research thrust for autonomous operation.

\subsection{Cross-Domain Collaboration: Robotics and Autonomy Lessons}
\label{subsec:crossdomain}
The journey toward an autonomous accelerator does not happen in isolation. It can benefit enormously from cross-disciplinary exchange, especially with the robotics and autonomous vehicle communities. Accelerators and robots may seem different, but both involve controlling complex physical systems in the real world under uncertainty, and both have seen a surge of AI-driven approaches. One open question is how techniques from autonomous robotics (which often involve navigation, manipulation, and agile decision-making in dynamic environments) can translate to accelerator control, and vice versa. For instance, robotics has developed robust methods for sim-to-real transfer, training control policies in simulation and then adapting them to work reliably on real hardware \cite{59_Pitkevich2024, 60_Chukwurah2024}. This is directly applicable to our accelerator digital twin approach (Section \ref{subsec:digitaltwin}). Tools like domain randomization, which expose the AI to many simulated variations, could help ensure that an RL agent trained on a simulated accelerator will not falter when faced with minor discrepancies on the real machine \cite{61_Peng2018}. Likewise, the safety frameworks in robotics (like emergency stop reflexes, multi-sensor fusion for obstacle avoidance) have analogues in accelerators (beam abort triggers, redundant readings for critical parameters). Exchanging ideas on these fronts will accelerate progress.

Conversely, advances made in the accelerator domain could inform other fields. The rigorous safety-first culture of accelerator operations and the integration of AI into a formal control system could provide a template for deploying AI in other safety-critical industries. The Osprey framework’s emphasis on transparent planning and constrained execution, for example, could be instructive for industrial robotics or nuclear plant AI where trust and verification are paramount. Additionally, an autonomous accelerator will likely need sophisticated coordination of many subsystems, akin to a factory of robots working in concert. Lessons from orchestrating accelerator AI agents might influence future multi-robot or smart factory control architectures. The synergy also extends to workforce and culture. Training a new generation of engineers fluent in both accelerator science and AI/robotics will be crucial. By actively collaborating and borrowing from the broader autonomy community, we can build on the foundation created by the self-driving car and robotics industries. This cross-domain approach will speed up development and help avoid pitfalls on the road to a truly autonomous accelerator.

\section{Challenges and Open Questions}

Despite the promise of AI-driven accelerators, significant challenges remain. Computationally, training and deploying sophisticated AI agents will require substantial hardware resources. The real-time inference demands of monitoring thousands of sensors, running digital twins in parallel, and coordinating multiple agents will necessitate significant investment in high-performance computing infrastructure. Data availability poses another hurdle. While accelerators generate vast amounts of operational data, certain failure modes and edge cases may be rare, making it difficult to train robust AI models for all scenarios. Transfer learning and simulation can help, but validation remains critical. How do we ensure an AI system trained primarily on simulated or historical data will respond appropriately to a novel fault condition never before encountered?

The human dimension is equally important. Transitioning to autonomous operation will require cultural change within the accelerator community and significant workforce development. A new generation of AI supervisors will need to be trained -- professionals fluent in both accelerator physics and AI systems who can oversee autonomous operations, interpret AI decision-making, intervene when necessary, and set strategic priorities. This represents both a challenge in developing appropriate training programs and an opportunity to attract diverse talent at the intersection of accelerator science and artificial intelligence.

Finally, regulatory and certification frameworks for AI in safety-critical roles are still evolving \cite{62_Terrier2023, 63_Ullrich2025, 64_Kusnirakova2023, 65_Ranjitsingh2025}. Demonstrating that an AI control system meets the same or higher safety standards as human operators will require rigorous testing, formal verification methods, and new approaches to system validation. The accelerator community must work proactively to establish appropriate standards and certification processes. What constitutes adequate testing for an AI system that will control a billion-dollar facility? How do we validate not just the AI’s performance under normal conditions, but its response to the full spectrum of possible failures? These are open questions that will require engagement between the AI and accelerator communities, safety experts, and regulatory bodies.

\section{Path Forward}
Realizing the vision of an autonomous, natively-AI particle accelerator will require sustained, coordinated effort across the accelerator physics, AI/ML, and controls engineering communities. The nine research thrusts outlined above are not independent. They form an interconnected ecosystem where progress in one area enables advances in others. For instance, robust digital twins (Section \ref{subsec:digitaltwin}) are prerequisites for safe reinforcement learning deployment (Section \ref{subsec:learning}), while multimodal state awareness (Sections \ref{subsec:knowledge} and \ref{subsec:multimodal}) underpins both agentic control architectures (Section \ref{subsec:agent}) and health monitoring systems (Section \ref{subsec:health}).

We envision a phased approach. Near-term efforts (2-5 years) should focus on maturing foundational capabilities. This includes expanding AI-assisted operations at existing facilities, developing high-fidelity digital twins for current machines, and establishing safety frameworks and transparency standards. These near-term efforts will move us from the current AI-assisted stage toward AI-augmented operations. Mid-term goals (5-10 years) involve progressively increasing AI autonomy, transitioning fully into AI-augmented operations with agents handling increasingly complex multi-subsystem tasks under human supervision. The long-term vision (10+ years) targets truly autonomous operation in the AI-autonomous stage, with new facilities designed from inception as AI-native systems where human operators serve primarily supervisory and strategic roles.

This transformation will require significant investment in both technology development and workforce training. However, the potential returns are substantial. Higher facility uptime and performance, reduced operational costs, and the ability to operate next-generation machines of unprecedented complexity are all within reach. The path to autonomous accelerators is challenging, but the foundation is already being laid through current AI deployments. By treating autonomy as a first-class design goal rather than an afterthought, we can shape the future of particle accelerator infrastructure for decades to come.

\section*{Acknowledgments}
This work was supported by the U.S. Department of Energy, Office of Science, Office of Nuclear Physics under Contract No. DE-AC05-06OR23177.

\bibliographystyle{abbrvnat}
\bibliography{references}  






\end{document}